\documentclass[12pt]{article}
\begin{document}
\def\bfone{\relax{\rm 1\kern-.35em 1}}\let\shat=\hat
\def\hat{\widehat}

%
\def\cS{{\cal K}}
\def\IE{\relax{{\rm I\kern-.18em E}}}
\def\cE{{\cal E}}
\def\rt{{\cR^{(3)}}}
\def\IGam{\relax{{\rm I}\kern-.18em \Gamma}}
\def\IGa{\IA}
\def\cV{{\cal V}}
\def\Rt{{\cal R}^{(3)}}
\def\tft#1{\langle\langle\,#1\,\rangle\rangle}
\def\IA{\relax{\hbox{{\rm A}\kern-.82em {\rm A}}}}
\def\hata{{\shat\a}}
\def\hatb{{\shat\b}}
\def\hatA{{\shat A}}
\def\hatB{{\shat B}}
\def\bv{{\bf V}}
\def\Fb{\overline{F}}
\def\nablab{\overline{\nabla}}
\def\Ub{\overline{U}}
\def\Db{\overline{D}}
\def\zb{\overline{z}}
\def\eb{\overline{e}}
\def\fb{\overline{f}}
\def\tb{\overline{t}}
\def\Xb{\overline{X}}
\def\Vb{\overline{V}}
\def\Cb{\overline{C}}
\def\Sb{\overline{S}}
\def\delb{\overline{\del}}
\def\Gammab{\overline{\Gamma}}
\def\Ab{\overline{A}}
\def\Anh{A^{\rm nh}}
\def\alphab{\bar{\alpha}}
\def\cy{Calabi--Yau}
\def\cabg{C_{\alpha\beta\gamma}}
\def\B{\Sigma}
\def\Bh{\hat \Sigma}
\def\Kh{\hat{K}}
\def\Knh{{\cal K}}
\def\A{\Lambda}
\def\Ah{\hat \Lambda}
\def\R{\hat{R}}
\def\V{{V}}
\def\T{T}
\def\Gammah{\hat{\Gamma}}
\def\twot{$(2,2)$}
\def\K{K\"ahler}
\def\rat{({\theta_2 \over \theta_1})}
\def\lv{{\bf \omega}}
\def\w{w}
\def\CP{C\!P}
\def\o#1#2{{{#1}\over{#2}}}
\def\eq#1{(\ref{#1})}

\def\ib{{\bar \imath}}
\def\jb{{\bar \jmath}}
\def\Im{{\rm Im ~}}
\def\Re{{\rm Re ~}}
\def\IP{\relax{\rm I\kern-.18em P}}
\def\arccosh{{\rm arccosh ~}}
%
\font\cmss=cmss10 \font\cmsss=cmss10 at 7pt
\def\twomat#1#2#3#4{\left(\matrix{#1 & #2 \cr #3 & #4}\right)}
\def\inbar{\vrule height1.5ex width.4pt depth0pt}
\def\IC{\relax\,\hbox{$\inbar\kern-.3em{\rm C}$}}
\def\IG{\relax\,\hbox{$\inbar\kern-.3em{\rm G}$}}
\def\IB{\relax{\rm I\kern-.18em B}}
\def\ID{\relax{\rm I\kern-.18em D}}
\def\IL{\relax{\rm I\kern-.18em L}}
\def\IF{\relax{\rm I\kern-.18em F}}
\def\IH{\relax{\rm I\kern-.18em H}}
\def\II{\relax{\rm I\kern-.17em I}}
\def\IN{\relax{\rm I\kern-.18em N}}
\def\IP{\relax{\rm I\kern-.18em P}}
\def\IQ{\relax\,\hbox{$\inbar\kern-.3em{\rm Q}$}}
\def\bfzero{\relax\,\hbox{$\inbar\kern-.3em{\rm 0}$}}
\def\IR{\relax{\rm I\kern-.18em R}}
\def\ZZ{\relax\ifmmode\mathchoice
{\hbox{\cmss Z\kern-.4em Z}}{\hbox{\cmss Z\kern-.4em Z}}
{\lower.9pt\hbox{\cmsss Z\kern-.4em Z}}
{\lower1.2pt\hbox{\cmsss Z\kern-.4em Z}}\else{\cmss Z\kern-.4em
Z}\fi}
\def\IU{\relax\,\hbox{$\inbar\kern-.3em{\rm U}$}}
\def\bfone{\relax{\rm 1\kern-.35em 1}}
\def\dop{{\rm d}\hskip -1pt}
\def\real{{\rm Re}\hskip 1pt}
\def\trace{{\rm Tr}\hskip 1pt}
\def\ii{{\rm i}}
\def\diag{{\rm diag}}
\def\sch#1#2{\{#1;#2\}}

\newcommand{\ft}[2]{{\textstyle\frac{#1}{#2}}}
\newcommand{\QED}{{\hspace*{\fill}\rule{2mm}{2mm}\linebreak}}
\def\dop{{\rm d}\hskip -1pt}
\def\bfone{\relax{\rm 1\kern-.35em 1}}
\def\bfzero{\relax{\rm I\kern-.18em 0}}
\def\inbar{\vrule height1.5ex width.4pt depth0pt}
\def\IC{\relax\,\hbox{$\inbar\kern-.3em{\rm C}$}}
\def\ID{\relax{\rm I\kern-.18em D}}
\def\IF{\relax{\rm I\kern-.18em F}}
\def\IK{\relax{\rm I\kern-.18em K}}
\def\IH{\relax{\rm I\kern-.18em H}}
\def\II{\relax{\rm I\kern-.17em I}}
\def\IN{\relax{\rm I\kern-.18em N}}
\def\IP{\relax{\rm I\kern-.18em P}}
\def\IQ{\relax\,\hbox{$\inbar\kern-.3em{\rm Q}$}}
\def\IR{\relax{\rm I\kern-.18em R}}
\def\IG{\relax\,\hbox{$\inbar\kern-.3em{\rm G}$}}
\font\cmss=cmss10 \font\cmsss=cmss10 at 7pt
\def\ZZ{\relax\ifmmode\mathchoice
{\hbox{\cmss Z\kern-.4em Z}}{\hbox{\cmss Z\kern-.4em Z}}
{\lower.9pt\hbox{\cmsss Z\kern-.4em Z}}
{\lower1.2pt\hbox{\cmsss Z\kern-.4em Z}}\else{\cmss Z\kern-.4em
Z}\fi}
\def\i{\rm i} 
\def\a{\alpha} \def\b{\beta} \def\d{\delta}
\def\e{\epsilon} \def\c{\gamma}
\def\G{\Gamma} \def\l{\lambda}
\def\L{\Lambda} \def\s{\sigma}
\def\cA{{\cal A}} \def\cB{{\cal B}}
\def\cC{{\cal C}} \def\cD{{\cal D}}
\def\cF{{\cal F}} \def\cG{{\cal G}}
\def\cH{{\cal H}} \def\cI{{\cal I}}
\def\cJ{{\cal J}} \def\cK{{\cal K}}
\def\cL{{\cal L}} \def\cM{{\cal M}}
\def\cN{{\cal N}} \def\cO{{\cal O}}
\def\cP{{\cal P}} \def\cQ{{\cal Q}}
\def\cR{{\cal R}} \def\cV{{\cal V}}\def\cW{{\cal W}}
%
%
%
\def\crr{\crcr\noalign{\vskip {8.3333pt}}}
\def\tilde{\widetilde}
\def\bar{\overline}
\def\us#1{\underline{#1}}
\let\shat=\hat
\def\hat{\widehat}
\def\hyp{\vrule height 2.3pt width 2.5pt depth -1.5pt}
\def\square{\mbox{.08}{.08}}

\def\Coeff#1#2{{#1\over #2}}
\def\Coe#1.#2.{{#1\over #2}}
\def\coeff#1#2{\relax{\textstyle {#1 \over #2}}\displaystyle}
\def\coe#1.#2.{\relax{\textstyle {#1 \over #2}}\displaystyle}
\def\half{{1 \over 2}}
\def\shalf{\relax{\textstyle {1 \over 2}}\displaystyle}
\def\dag#1{#1\!\!\!/\,\,\,}
\def\to{\rightarrow}
\def\notin{\hbox{{$\in$}\kern-.51em\hbox{/}}}
\def\shdot{\!\cdot\!}
\def\ket#1{\,\big|\,#1\,\big>\,}
\def\bra#1{\,\big<\,#1\,\big|\,}
\def\equaltop#1{\mathrel{\mathop=^{#1}}}
\def\Trbel#1{\mathop{{\rm Tr}}_{#1}}
\def\inserteq#1{\noalign{\vskip-.2truecm\hbox{#1\hfil}
\vskip-.2cm}}
\def\attac#1{\Bigl\vert
{\phantom{X}\atop{{\rm\scriptstyle #1}}\phantom{X}}}
\def\exx#1{e^{{\displaystyle #1}}}
\def\del{\partial}
\def\delbar{\bar\partial}
\def\nex#1{$N\!=\!#1$}
\def\dex#1{$d\!=\!#1$}
\def\cex#1{$c\!=\!#1$}
\def\eg{{\it e.g.}} \def\ie{{\it i.e.}}
%
\def\cS{{\cal K}}
\def\IE{\relax{{\rm I\kern-.18em E}}}
\def\cE{{\cal E}}
\def\rt{{\cR^{(3)}}}
\def\IGam{\relax{{\rm I}\kern-.18em \Gamma}}
\def\IGa{\IA}
\def\cV{{\cal V}}
\def\Rt{{\cal R}^{(3)}}
\def\tft#1{\langle\langle\,#1\,\rangle\rangle}
\def\IA{\relax{\hbox{{\rm A}\kern-.82em {\rm A}}}}
\def\hata{{\shat\a}}
\def\hatb{{\shat\b}}
\def\hatA{{\shat A}}
\def\hatB{{\shat B}}
\def\bv{{\bf V}}
\def\Fb{\overline{F}}
\def\nablab{\overline{\nabla}}
\def\Ub{\overline{U}}
\def\Db{\overline{D}}
\def\zb{\overline{z}}
\def\eb{\overline{e}}
\def\fb{\overline{f}}
\def\tb{\overline{t}}
\def\Xb{\overline{X}}
\def\Vb{\overline{V}}
\def\Cb{\overline{C}}
\def\Sb{\overline{S}}
\def\delb{\overline{\del}}
\def\Gammab{\overline{\Gamma}}
\def\Ab{\overline{A}}
\def\Anh{A^{\rm nh}}
\def\alphab{\bar{\alpha}}
\def\cy{Calabi--Yau}
\def\cabg{C_{\alpha\beta\gamma}}
\def\B{\Sigma}
\def\Bh{\hat \Sigma}
\def\Kh{\hat{K}}
\def\Knh{{\cal K}}
\def\A{\Lambda}
\def\Ah{\hat \Lambda}
\def\R{\hat{R}}
\def\V{{V}}
\def\T{T}
\def\Gammah{\hat{\Gamma}}
\def\twot{$(2,2)$}
\def\K{K\"ahler}
\def\rat{({\theta_2 \over \theta_1})}
\def\lv{{\bf \omega}}
\def\w{w}
\def\CP{C\!P}
\def\o#1#2{{{#1}\over{#2}}}
\def\eq#1{(\ref{#1})}
\newcommand{\be}{\begin{equation}}
\newcommand{\ee}{\end{equation}}
\newcommand{\ba}{\begin{eqnarray}}
\newcommand{\ea}{\end{eqnarray}}
\newtheorem{definizione}{Definition}[section]
\newcommand{\bd}{\begin{definizione}}
\newcommand{\ed}{\end{definizione}}
\newtheorem{teorema}{Theorem}[section]
\newcommand{\bth}{\begin{teorema}}
\newcommand{\eth}{\end{teorema}}
\newtheorem{lemma}{Lemma}[section]
\newcommand{\blem}{\begin{lemma}}
\newcommand{\elem}{\end{lemma}}
\newcommand{\brr}{\begin{array}}
\newcommand{\err}{\end{array}}
\newcommand{\nn}{\nonumber}
\newtheorem{corollario}{Corollary}[section]
\newcommand{\bcorol}{\begin{corollario}}
\newcommand{\ecorol}{\end{corollario}}
\def\twomat#1#2#3#4{\left(\begin{array}{cc}
 {#1}&{#2}\\ {#3}&{#4}\\
\end{array}
\right)}
\def\twovec#1#2{\left(\begin{array}{c}
{#1}\\ {#2}\\
\end{array}
\right)}
\thispagestyle{empty}
\begin{titlepage}
\thispagestyle{empty}
\begin{flushright}
CERN-TH/98-234\\
July 1998\\
\end{flushright}
\vskip 2.cm
\begin{center}
{\Large\bf ``Non chiral'' primary superfields in the $AdS_{d+1} / CFT_d$
correspondence
\footnote{Work
 supported in part by EEC under TMR contract ERBFMRX-CT96-0045 (LNF Frascati
 and Politecnico di Torino), Angelo Della Riccia fellowship
 and by DOE grant DE-FG03-91ER40662}}
\vskip 2.cm
{ Laura Andrianopoli and Sergio Ferrara}\\
{\it  CERN Theoretical Division, CH 1211 Geneva 23, Switzerland.}
\end{center}

\begin{abstract}
We consider some long multiplets describing bulk massive excitations 
of M-theory two-branes and IIB string three-branes which correspond to
``non chiral'' primary operators of the boundary $OSp(8/4)$ and $SU(2,2/4)$
superconformal field theories.

Examples of such multiplets are the ``radial'' modes on the branes, including
 up to spin 4 excitations, which may be then considered as prototypes 
of states which are not described by the K-K spectrum of the corresponding
 supergravity theories on $AdS_4 \times S_7$ and $AdS_5 \times S_5$
 respectively.
\end{abstract}
\end{titlepage}
\section{Introduction}
The conjectured $AdS_{d+1} / CFT_d$ correspondence \cite{jm, gkp, w1, greci}
 naturally associates the
 spectrum of the K--K excitations of the supergravity theory 
\cite{van, krw, gm} describing the
 horizon geometry  of M-theory and string theory branes \cite{gt, gib, fre}
 to the  spectrum of
 ``chiral'' primary conformal operators of the boundary conformal field theory
\cite{ho, ff, w1, ffz, aoy},
 related to the brane world-volume theory \cite{kal}.
A detailed analysis of this correspondence has been carried out for all 
maximally supersymmetric theories \cite{w1, ffz, aoy}, corresponding to 16
 (Poincar\'e)
 supersymmetries on the world-volume theory, and also for many examples with 
reduced supersymmetry, corresponding to matter-coupled anti de Sitter
 supergravity in the bulk theory.
\par
A particular property of these states is that their $AdS$ mass is quantized 
and this corresponds to the absence of anomalous dimensions of the 
corresponding ``operators'' describing the very same representations when the 
same superalgebra is realized as conformal field theory on the anti de Sitter 
boundary \cite{w1}.
On the other hand the $CFT_d$ naturally contains operators which have the 
same quantum numbers of string or M-theory excitations not present in the K-K
spectrum of the corresponding $AdS_{d+1}$ supergravity theory
 \cite{w1, gkp, irk}.
These states have ``anomalous dimensions'', which in type IIB string
compactified on $AdS_5 \times S_5$ grow as $(g^2 N)^{1/4}$, as a consequence 
of the relation between $AdS$ masses and conformal dimensions of the alluded 
correspondence \cite{gkp}.
An immediate consequence of this fact is that these excitations are  described
by ``non chiral'' primary superfields representations which, in $AdS_5$, 
describe ``massive'' string excitations in ``long multiplets'' of the
 $SU(2,2/4)$ superalgebra \cite{fz}.
\par
In the present paper we describe the excitations corresponding to these
 multiplets and their degeneracy, which is a consequence of the underlying 
extended superalgebra.
\par
Without loss of generality we will describe the simplest of these long
multiplets in M-theory on $AdS_4 \times S_7$ and IIB string theory on $AdS_5
 \times S_5$.
This multiplet is associated to the radial mode of the two- and three-brane
 respectively, i.e. it is the supermultiplet whose (lowest dimensional)
 component is 
 $ Tr (X_T X_T) $,
where $X_T$ are the ``coordinates'' transverse to the brane.
Here the trace is taken over some Lie Algebra, present for the case of N 
branes,
which however does not play any role in our discussion.
It turns out that this multiplet is actually the ``simplest'' long
 supermultiplet
 of the corresponding superalgebra and it contains $2^{16}$ states with spin 
range from 0 up to 4.
\par
The paper is organized as follows:\\
In section II we will review some properties of massive extended superfields.
In section III we will describe the $d=3$ case, corresponding to the M-theory
2-brane.
This case is technically simpler even if the corresponding superconformal field
 theory exists only as the infrared limit of the gauge theory on the brane and
 is not known.
In section IV we will then consider the case of IIB theory on $AdS_5 \times S_5$
and describe the simplest long multiplet of the $SU(2,2/4)$ superalgebra, 
sustaining ``anomalous dimensions'' for the primary conformal operators.
This multiplet contains the same number of states as the $d=3$ example even if
 the interpretation is different.
We will give the quantum numbers and degeneracy of the bulk ``excitations'' 
corresponding to these multiplets, with regard to spherical harmonics on $S_5$.
\par
The paper will end with some conclusions and outlooks.
\section{Long multiplets in extended supersymmetry}
In this section we review some properties of extended superfields in $d$-dimensional 
Minkowski space $M_d$, for the cases $d=3$, $d=4$.
\par
In the case of N-extended supersymmetry, the $d=3,4$ algebras have a R-symmetry
$O(N)$ and $U(N)$ respectively ($SU(4)$ for $N=4$).
The theories with maximal (conformal) supersymmetry correspond to $N=8$ at $d=3$ 
and $N=4$ at $d=4$, sustaining an algebra with 16 Poincar\'e supersymmetries.
Poincar\'e superfields can be enlarged to conformal superfields, in the sense 
that a Poincar\'e superfield, through the method of induced representations, 
can induce a representation of the full superconformal algebra
 \cite{bdw, hst, hw, af}.
If the superfield is unrestricted, it can carry ``anomalous dimensions''.
However, if the superfield is restricted, such as chiral superfields are, 
then the constraints are compatible with conformal invariance only if the
``conformal dimension'' is ``quantized''.
This is the origin of the quantized spectrum of ``chiral primary superfields''
which describe the K--K excitations of supergravity on $AdS_{d+1}$ in the
 $CFT_d$
correspondence.
Examples of such superfields, with quantized conformal dimensions, are given
 in the $N=1$, $d=4$ theory by the chiral multiplets: 
$ \bar D_{\dot\alpha} S_{\{\alpha_1 \cdots \alpha_n\}}=0$,
where $S$ is a $(J,0)$ representation of $SL(2,C)$, or by the ``current
 multiplets'' $J_{\alpha\dot\alpha}, J$, such that \cite{ffz}: $
  D^\alpha J_{\alpha\dot\alpha}= D^{\dot\alpha} J_{\alpha\dot\alpha}=0$,
$ D^2 J = \bar D^2 J =0$.
\begin{sloppypar}
These superfields describe respectively ``hypermultiplets'', ``gravity multiplet'' and ``vector
 multiplets''  in the bulk $N=2$ supergravity theory on $AdS_5$,
 with an 
underlying gauge algebra $U(2,2/1) \times G_f$, where $G_f$ is some flavour
 symmetry
of the boundary $SCFT_4$.
The conformal dimension $\ell$ of these superfields is $\ell = q_J$ ($q_J$ is 
the $U(1)$ charge of the $U(2,2/1)$ algebra for the lowest component $S|_{\theta =0}$), 
$\ell =3$ and $\ell = 2$ respectively \cite{fz2}.
\\
Note that, for ``short multiplets'', the number of components 
is always $2^2 \times r$, with $r$ some spin representation (hypermultiplets
 have $r=2$,
the gravity multiplet $r=4$, vector multiplets $r=2$).
\\
In $N=1$ an unconstrained superfield is a real scalar superfield which has 
$2^4 =16$ components, giving $4(0,0)$, $1 (\frac{1}{2} ,\frac{1}{2} )$, $2[(\frac{1}{2} ,0) + 
(0,\frac{1}{2} )]$ states.
These states are easily seen to describe a ``massive'' $N=2$ long multiplet in
 the $AdS_5$ correspondence.\\
Since in the $AdS /CFT$ correspondence the conformal dimension $\ell$ and the
 $SL(2,C )$ quantum numbers $(J_1,J_2)$ of a given operator on the boundary 
are mapped into the quantum numbers $(E_0,J_1,J_2)$ of the maximal compact
 subgroup $O(2) \times SU(2)\times SU(2)$ of $O(4,2)$ acting on states in the
 bulk, we will often interchange the notations.
\end{sloppypar}
\par
Let us now consider a generic $N$-extended theory at $d=4$.
The R-symmetry is $U(N)$ ($SU(4)$ for $N=4$).
If the conformal superfield is ``short'' (or chiral) then the superfield will
 contain $2N$ 
$\theta$'s (instead of $4N$) and the $\theta$ expansion will give $2^{2N}$
 states.
These representations have been described elsewhere and will not be repeated
 here \cite{gmz}.\\
Supersingletons correspond to ``ultrashort representations'' with multiplicity
 $2^N$
(up to CPT doubling when required). These representations have no particle
 interpretation on $AdS_5$ \cite{fl1, fl2, fl3, gm, gmz}.
\par
For long superfields the $\theta$ expansion gives instead $2^{4N}$ states. 
These states can be thought as a basis for a representation of the Clifford
 algebra of the orthogonal group $O(8N)$.\\
The left and right representations of $O(8N)$ correspond to bosons and 
fermions, i.e. to even and odd powers  of $\theta$'s \cite{fs}.
\par
To classify the states with respect to $AdS_5$ quantum numbers we have simply
 to decompose the rep. of the Clifford algebra, that is the spinor
 representation of $O(8N)$ (long multiplets)
with respect to the $O(4)$ spin symmetry and $U(N)$ R-symmetry.
\\
In the case of $O(8N)$ the decomposition is as follows \cite{fuchs}:
$  8N \to (4,2N) $,
under $Sp(4) \times Sp(2N)$, with the further embedding:
\begin{eqnarray}
  Sp(4) \to O(4) && (4 \to (\frac{1}{2} ,0) + (0,\frac{1}{2} ))\\
Sp(2N) \to U(N) && (2N \to N + \bar N). 
\end{eqnarray}
\\
It follows that each $O(4)$ rep. completes a $O(5)$ rep. and each $SU(4)$ rep. 
completes a $Sp(8)$ rep.
States are therefore naturally classified under $O(5) \times Sp(8)$.

In $d=3$ the story is very similar, but even simpler.
In this case the spin subgroup of $AdS_4$ is simply $SU(2)$ and the R-symmetry
 of $SCFT_3$ is $O(N)$, with 2-component Majorana spinors.
Maximal conformal supersymmetry corresponds to $N=8$, this being related to
 $N=8$ supergravity on $AdS_4$.
Short representations of the $OSp(8/4)$ superalgebra both on $AdS_4$ and
 $\partial AdS_4$
 have been described elsewhere.
The simplest of them is the $N=8$ graviton multiplet which is obtained 
tensoring the singleton 
representation $(\phi_s , \psi_c)$ where $\phi_s , \psi_c$ are a massless 
conformal
scalar and spin-$\frac{1}{2}$ fields in the two 8 dimensional  spinor 
representations $8_s$, $8_c$
 of $O(8)$.
This multiplet contains $2^{8}$ states corresponding to the $N=8$ supergravity
 multiplet \cite{aoy}.
\par
Long supermultiplets correspond to massive supermultiplets in $AdS_4$ which do
 not correspond to ``chiral'' primary superfields on the boundary.
The $\theta$ expansion will involve 16 $\theta$'s and contains $2^{16}$ states
 with spin range $s=0, \cdots , \frac{N}{2} = 4 $.\\
To classify the degeneracy of these states with respect to the spin group
 $SU(2)$ and the R-symmetry $O(N)$ we decompose $O(4N)$ as follows:
$  4N \to (2,2N) $,
under $O(4N) \to SU(2) \times Sp(2N)$, with the further decomposition:  
$  2N \to (2,N)$, 
under $Sp(2N) \to SU(2) \times O(N)$ .
From the above we then note that massive long multiplets have a high 
degeneracy.
For  $N=8$, states of a given spin are classified by $Sp(16)$ reps.
Then $Sp(16)$ representations are naturally decomposed under the R-symmetry
 group $O(8)$ and
an ``internal'' $SU_I(2)$ symmetry which gives the degeneracy of states in the
 same
 spin $SU(2)$ and R-symmetry $O(8)$ representation.
\section{Non chiral primary operators in the
 $AdS_4 / CFT_3$ correspondence:
 M-theory on $AdS_4 \times S_7$}
In M-theory 2-branes the fundamental conformal multiplet can be taken to be 
the supersingleton representation of $OSp(8/4)$.
\par
By a choice of triality basis, this multiplet contains  a 3-$d$ boson and a 
3-$d$ (Majorana) fermion in the $8_s$, $8_c$ spinor reps. of $O(8)$
 respectively.
This choice assigns the spinor charges to the vector rep. of $O(8)$ as in
 standard $N=8$ supergravity in $AdS_4$.
\par
In the dynamics of N M-branes, these multiplets are assumed to be Lie algebra
 valued in some group $G$ which depends on the N branes gauge dynamics
\cite{aoy}.
\par
The ``chiral primary'' operators are obtained by superfields whose lowest
 component is:
$  Tr(8_{s_1}\cdots 8_{s_p}) - \mbox{traces}$,
i.e. gauge singlets in the ``totally symmetric'' irreducible multi spinor
 representation of $O(8)$.
\par 
The 2-branes coordinates and their superpartners are the supersingleton octets
\cite{gt}.
It is obvious that we can make bilinears (or multilinears) in singletons
which do not give short representations with spin $s\le 2$.
In fact, by suitable multiplication, in the product of two (massless)
 supersingletons we may get arbitrarily high spin massless representations with
 the following structure \cite{ss}:
 \begin{equation}
   \underbrace{2s-2}_1 ,\underbrace{2s-{3}/{2}}_{8} , 
\underbrace{2s-1}_{28},\cdots ,
\underbrace{2s}_{35^++35^-},\cdots , 
 \underbrace{2s+1}_{28}, \underbrace{2s+{3}/{2}}_{8},
 \underbrace{2s+2}_1
\label{mult}
 \end{equation}
Conserved currents of spin $s$ on the boundary of $AdS_4$ give two physical 
degrees of freedom (of helicity $\pm s$) in the bulk, as appropriate for
 massless particles \cite{ff}.
Therefore the above multiplet contains $2\times 2^8$ states (unless $s=0$).\\
The  multiplet (\ref{mult}), for $s=0$, is the graviton multiplet with scalars in 
the $35^+$, $35^-$ (in this case the degrees of freedom are $2^8$).
This corresponds to a superfield starting with the $35$ of $8_s \times 8_s
 |_S=1+35$.
\par
The spin 4 multiplet ($s=1$) corresponds instead to a massless superfield,
 starting with the singlet in $8_s \times 8_s|_1$.
This is what we call the radial mode on the brane.
In the next section we will see that a similar multiplet exists for the 3-brane
 of type IIB
string on $AdS_5 \times S_5$.
\par
We now consider the simplest superfield which corresponds to a long multiplet
 of the $OSp(8/4)$
algebra.
\\
In a hypothetical 3-$d$ interacting theory of singletons, we may think the
 multiplet to  be
obtained by a scalar unrestricted $N=8$ superfield whose lowest component is:
$Tr(\phi_s\phi_s)$,
where however now we do not assume $\phi_s$ (and $\psi_c$) to be massless.
\\
As a result we expect to obtain a massive spin 4 multiplet with $2^{16}$
 components 
classified in different representations of the spin $SU(2)$ and of the
 R-symmetry $O(8)$.
\\
In this case the decomposition of the Clifford algebra of $O(32)$ with respect
 to $SU(2)\times Sp(16)$
is straigthforward \cite{fs} because it is the same as the one which occurs
 for long Poincar\'e 
multiplets of $N=8$ Poincar\'e supergravity in $D=4$.
\par
The result is that the representation of the Clifford algebra of $O(32)$
 decomposes in a
 direct sum of irreducible (antisymmetric traceless) reps. of $Sp(16)$, where 
the $k$-fold 
antisymmetric component corresponds to spin $s=4-\frac{k}{2}$ ($k=0,\cdots ,8$)
 each representation occurring with multiplicity one.
\\
The R-symmetry content of each of these $Sp(16)$ reps. can be further analysed
 by decomposing
$Sp(16)$ with respect to the maximal subgroup $SU_I(2) \times O(8)$ \cite{sla},
 where $SU_I(2)$ is an internal
spin which gives the degeneracy of each $O(8)$ rep. for a given (space--time)
spin $s$.
\\
This multiplet is an obvious candidate for M-theory massive excitations
 obtained from 
the $AdS_4 /CFT_3$ correspondence.\\
Note that since the spectrum of M-theory on $AdS_4 \times S_7$ is unknown, we
 cannot be sure that
such a multiplet really occurs in the spectrum, but what we mean is that any
 long massive
 multiplet will have a similar structure.
Indeed the multiplet considered before is the smallest long massive 
representation in
$AdS_4$, as much as the same as the graviton multiplet is the smallest short 
rep. in $AdS_4$.\\
In tables \ref{3dclass}, \ref{o8rep} we report the massive excitations contained in
 the long multiplet, 
corresponding to a scalar unrestricted $N=8$ superfield on $\partial AdS_4$, and the related
R-symmetry representations.
\\
Note that, for a given spin $s$, the internal spin $SU_I(2)$ classifies the
 conformal dimensions
of operators in a given $O(8)$ representation according to the formula:
\begin{equation}
  \label{quantumnumber3}
  \ell_{J_3,J,s}=\ell+J_3+4\, ;\,\, -J\le J_3 \le +J \, ;\, \, 0 \le J \le 4-s
\end{equation}
where we denoted with $\ell$ the conformal dimension of the lowest component 
of the superfield.\\
From the spectrum we may just notice that the only $O(8)$ singlets are a spin
 4 state (with conformal 
dimension $\ell + 4$), which is a scalar under $SU_I(2)$, and nine scalar
 states 
(with conformal dimensions $\ell , \ell +1,\cdots ,\ell +8$), which have spin 4 under $SU_I(2)$.
\par
In the free field theory limit ($\ell =1$), where the singletons are massless, this multiplet shrinks
to a massless representation of the $OSp(8/4)$ superalgebra and only two scalar states remain, 
in $Tr \phi_s\phi_s$ with $\ell =1$ and in $Tr \psi_c\psi_c$ with $\ell =2$.
\par
Also the multiplicity of the higher spin states is just one for each
antisymmetric rep. of $O(N)$, according to (\ref{mult}).

\section{``Non chiral'' primary operators in the 
$AdS_5 / CFT_4$ correspondence:
 IIB string theory on $AdS_5 \times S_5$}
\begin{sloppypar}
In $d=4$, the $AdS/CFT$ correspondence is much more interesting because it
relates 4-$d$ superconformal invariant Yang--Mills theory to extended supergravity
in $AdS_5$ with an underlying superalgebra $U(2,2/N)$.
\end{sloppypar}
\par
The maximal case, corresponding to IIB string theory compactified on $AdS_5 
\times S_5$,
 corresponds to $d=4$, $N=4$ Yang--Mills theories, which are conformal
 invariant for arbitrary values
of the Yang--Mills coupling.
\par
The K--K spectrum of ten dimensional IIB supergravity on $AdS_5 \times S_5$
 has been considered
 in ref. \cite{krw, gm}, and its correspondence with the ``twisted chiral''
 primary superfields of
the $N=4$ superconformal algebra has been shown in detail in ref. 
\cite{fz, af}.
\\
The ``chiral'' primary superfields correspond to short multiplets whose lowest
 component is \cite{hw}: $  Tr(\phi_{\{\ell_1 \cdots} \phi_{\ell_p\}})-\mbox{traces}$,
where $\phi_\ell$ is a $SU(N)$ Lie-algebra valued ``transverse'' coordinate and
 the $(0,p,0)$ $SU(4)$ rep.
is singled out.
\par
In ref. \cite{gkp} Gubser, Klebanov and Polyakov have pointed out that a
``Yang--Mills'' composite operator which couples to string states exists
 such as: $
 Tr(F_{\mu_1 \nu_1}\nabla_{\alpha_1}\cdots \nabla_{\alpha_n}
 F_{\mu_2 \nu_2}) 
 $,
where $F_{\mu\nu}$ is the Yang--Mills field strength.
These operators, being related to string massive states,  have anomalous dimensions
$\delta$ which, through the $AdS_5 /CFT_4$ correspondence,  are predicted to grow, at
strong coupling $g^2 N$ large, as $\delta = (g^2N)^{1/4}$.
\par
In the context of $N=4$ superconformal symmetry, this necessarily implies that
these operators are members of ``long multiplets'', naturally associated to
 ``excitations''
 in $AdS_5$ which do not correspond to ``short reps'' of the $SU(2,2/4)$
 algebra.
\par
In this section we analyze in detail the simplest of these excitations,
 contained in the product
of two singleton representations, namely the superfield whose lowest component
 is the 
``radial mode'' on the brane: $
  \Phi = Tr(\phi_\ell \phi^\ell)$.
This composite scalar is expected to have anomalous dimension (growing as 
$(g^2N)^{1/4}$ in
the strong coupling regime) and contains excitations on $AdS_5 \times S_5$ up
 to spin 4 \cite{hst}.
\\
This multiplet is the $N=4$ version of the so-called Konishi multiplet
 \cite{kk}, which in $N=1$ notation is a real unrestricted superfield 
satisfying the relation: $
  \bar D \bar D \Sigma = W$,
where $
 \Sigma =S_i e^{gV} \bar S_i \, , \quad 
W = g f_{\Lambda\Sigma\Delta}\epsilon^{ijk}S_i^\Lambda S_j^\Sigma S_k^\Delta$,
$i=1,2,3$, and $\Lambda ,\Sigma ,\Delta  \in Adj SU(N)$.\\
In the free field theory limit this multiplet corresponds to a short conserved
 current multiplet
 with $\bar D \bar D \Sigma = DD\Sigma =0$ and $\ell = 2$.
It corresponds to a massless vector multiplet in $AdS_5$ according to the
discussion of section II.
However in the non abelian gauge theory the $\Sigma$ multiplet becomes ``long''
 and corresponds
 to a ``massive vector multiplet'' in $AdS_5$. 
Some properties of this massive multiplet, containing $2^{16}$ states, 
were given in ref. \cite{fz}.
Here we give the complete spectrum of the $AdS_5$ representations and their degeneracy.
\par
Let us first notice that for an orthogonal group $O(4NM)$ there is always a maximal
subalgebra $Sp(2N)\times Sp(2M)$, according to the embedding \cite{fuchs}: 
$  4NM \to (2N,2M)$
of the vector representation.
A particular case of this decomposition ($N=1, M=8$) was used in the previous section.
\\
On the $4d$ boundary of $AdS_5$ the relevant decomposition is with respect to a group which 
contains the space--time spin $O(4)$ and the R-symmetry $SU(4)$ of the corresponding $SU(2,2/4)$
superalgebra.
Then we see that the appropriate decomposition corresponds to $N=2$, $M=4$ i.e.: $
  O(32) \to Sp(4) \times Sp(8)$, $ 32 \to (4,8)$,
with the further decomposition:
\begin{eqnarray}
  4 \to (\frac{1}{2} , 0) + (0, \frac{1}{2} ) \, & \mbox{ under } & Sp(4) \to SU(2) \times SU(2) \nonumber\\
 8 \to 4+\bar 4  \, & \mbox{ under } & Sp(8) \to SU(4)\times U(1)
\end{eqnarray}

Having defined the embedding it is now straigthforward to analyze the $O(32)$
 Clifford algebra in terms of $Sp(4)\times Sp(8)$ representations.
They are given by the Young tableaux exhibited in table \ref{4dclass} 
($U(1)$ charges have not been given here).
\par
The further decomposition in $SU(2) \times SU(2) \times SU(4)$ representations
is given in tables \ref{sp4dec}, \ref{sp8dec}, \ref{su4rep}.
\\
As in $d=3$ there is a sense to make the massless limit of this massive
 representation.
This corresponds to consider the product $Tr(\phi_\ell \phi_\ell )$ where
 $\phi_\ell $ is a free field \cite{gmz}.
In this case one obtains a massless multiplet in $AdS_5$ up to spin 4, which
 contains 
$2^8 (2(J_L+J_R)+1)$ states with $J_L+J_R=2$.
The spectrum of the massless multiplets is given by Table 2 of ref. \cite{hst} 
where all $(J_L , J_R )$ conformal fields, with $J_L , J_R \ge \frac{1}{2}$, are
conserved currents on the boundary (and have therefore $E_0=2+J_L+J_R$).
This is the same spectrum as obtained in Table 12 of ref. \cite{gmz}.\\
As before we may easily identify states which are s-waves with respect to an hypothetical
partial wave  analysis on $S_5$.
\par
The $SU(4)$ singlets occur in the 1 (1), 36 (1), 42 (2), 308 (1), 825 (1) and
 594 (3) of $Sp(8)$.
This gives rise to 7 scalar singlets, 4 $(\frac{1}{2} , \frac{1}{2} )$ vectors, 5 (1,1)
 tensors, 2
 $(\frac{3}{2},\frac{3}{2})$ tensors, 2 $((2,0),(0,2),(\frac{3}{2},\frac{1}{2} ),
 (\frac{1}{2} ,\frac{3}{2} ))$ tensors and 1 (2,2) tensor. 
The $SU(4)$ s-waves have the following value of the energy:
\begin{eqnarray}
  (0,0) && E_0=\ell , \ell +2 ,\ell +4 , \ell +4 , \ell + 4 , \ell + 6 , \ell + 8 \nonumber\\
(\frac{1}{2} , \frac{1}{2} ) && E_0 = \ell +1 ,\ell +3 ,\ell +5 ,\ell +7\nonumber\\
(1,1) && E_0 =\ell +2 ,\ell +4 ,\ell +4, \ell +4,\ell +6 \nonumber\\
(\frac{3}{2} , \frac{3}{2}) && E_0 =\ell +3,\ell +5  \nonumber\\
(2,0),(0,2) &&  E_0 = \ell +2,  \ell +6\nonumber\\
(\frac{3}{2},\frac{1}{2} ), (\frac{1}{2} ,\frac{3}{2} ) && E_0 = \ell +3,  \ell +5\nonumber\\
(2,2) && E_0 = \ell +4
\end{eqnarray}
\section{Concluding remarks}
In this paper we have analyzed the structure of long multiplets,
in the spirit of the $AdS /CFT$ correspondence, for the maximally 
supersymmetric theories  corresponding to $d=3$, $d=4$  SCFT's with 16
 (Poincar\'e) spinorial charges.
As prototypes of these multiplets we considered the ``radial mode''
of the corresponding ($d-1$)-brane, which contains up to spin 4 excitations
and cannot therefore be described by supergravity on $AdS_{d+1} \times 
S_{D-d-1}$.
\par 
Multiplets with these properties are expected to describe M-theory and 
string theory higher spin excitations.
They can sustain anomalous dimensions and correpond to long
massive multiplets on the $AdS$ bulk. 
It is in principle possible to construct such kind of superconformal
 multiplets for arbitrarily high spin.
The counterpart of this in a non interacting conformal field theory is that 
it is possible to construct superconformal fields of arbitrary spin 
corresponding to conserved currents on the boundary, describing therefore
massless fields in the bulk theory. 
It is likely that such operators for $d=4$, $N=4$ Yang--Mills theory give rise,
in the corresponding theory on $AdS_5$, to stringy corrections \cite{bg, bgkr}
 in the 
scattering amplitudes for massless states.
The latter can be computed from OPE techniques on the boundary conformal field
 theory \cite{free, fz, liu}
 \section*{Acknowledgements}
\begin{sloppypar}
We acknowledge stimulating discussions with R. D'Auria, M. G\"unaydin,
R. Stora and A. Zaffaroni.
\end{sloppypar}

\section*{Tables}
\begin{table}[hb]
\caption{Massive excitations on $AdS_4$;
 $O(32) \to SU(2)\times Sp(16) \to SU(2) \times SU(2)_I\times O(8)$ classification }
\label{3dclass}
    \begin{tabular}{lll}
\hline
spin & $Sp(16)$ & $(SU(2)_I , O(8))$ \\
\hline 
4 & 1 &$ (1,1)$ \\
$\frac{7}{2}$ & 16 & $ (2 , 8)$ \\
3 & 119 & $(3 , 28 ) + (1, 35)$ \\
$\frac{5}{2}$ & 544 & $(4, 56)+(2 , 160)$ \\
2 & 1700 & $(5,70)+(3,350)+(1,300)$\\
$\frac{3}{2}$ & 3808 & $(6, 56)+(4, 448)+(2 , 840)$\\
1 & 6188 & $(7,28) +(5,350)+(3,1134)+(1,840')$\\
$\frac{1}{2} $& 7072 & $(8,8)+(6,160)+(4,840)+(2 ,1344)$\\
0 & 4862 & $(9,1)+(7,35)+(5,300)+(3,840')+(1,588)$\\
\hline
    \end{tabular}
 \end{table}

\setlength{\unitlength}{1mm}
\begin{table}[t]
 \caption{$O(8)$ reps of interest}
\label{o8rep}
    \begin{tabular}{lcl}
\hline
rep & Young tableau  & Dynkin label \\
\hline
1 & \bfone
  &  (0,0,0,0) \\
8 & \begin{picture}(8,8)
\put(0,0){\framebox (2,2)}
\end{picture} & (1,0,0,0) \\
28 &\begin{picture}(8,8)
\multiput(0,0)(0,-2){2}{\framebox (2,2)}
\end{picture} &  (0,1,0,0)) \\
35 &\begin{picture}(8,8)
\multiput(0,0)(2,0){2}{\framebox (2,2)}
\end{picture} &  (2,0,0,0)) \\
56 &\begin{picture}(8,8)
\multiput(0,0)(0,-2){3}{\framebox (2,2)}
\end{picture} &  (0,0,1,1) \\
160 &\begin{picture}(8,8)
\multiput(0,0)(0,-2){2}{\framebox (2,2)}
\put(2,0){\framebox (2,2)}
\end{picture} &  (1,1,0,0) \\
$70=35^+ +35^-$ &\begin{picture}(8,8)
\multiput(0,0)(0,-2){4}{\framebox (2,2)}
\end{picture} &  (0,0,0,2)+(0,0,2,0) \\
350 &\begin{picture}(8,8)
\multiput(0,0)(0,-2){3}{\framebox (2,2)}
\put(2,0){\framebox (2,2)}
\end{picture} &  (1,0,1,1) \\
300 & \begin{picture}(8,8)
\multiput(0,0)(2,0){2}{\framebox (2,2)}
\multiput(0,-2)(2,0){2}{\framebox (2,2)}
\end{picture} & (0,2,0,0) \\
$448=224^++224^-$ &\begin{picture}(8,8)
\multiput(0,0)(0,-2){4}{\framebox (2,2)}
\put(2,0){\framebox (2,2)}
\end{picture} &  (1,0,0,2)+(1,0,2,0) \\
840 & \begin{picture}(8,8)
\multiput(0,0)(0,-2){3}{\framebox (2,2)}
\multiput(2,0)(0,-2){2}{\framebox (2,2)}
\end{picture} & (0,1,1,1) \\
840' & \begin{picture}(8,8)

\multiput(0,0)(0,-2){3}{\framebox (2,2)}
\multiput(2,0)(0,-2){3}{\framebox (2,2)}
\end{picture} & (0,0,2,2) \\
$1134=567^++567^-$ & \begin{picture}(8,8)
\multiput(0,0)(0,-2){4}{\framebox (2,2)}
\multiput(2,0)(0,-2){2}{\framebox (2,2)}
\end{picture} & (0,1,0,2)+(0,1,2,0) \\
$1344=672^++672^-$ & \begin{picture}(8,8)
\multiput(0,0)(0,-2){4}{\framebox (2,2)}
\multiput(2,0)(0,-2){3}{\framebox (2,2)}
\end{picture} & (0,0,1,3)+(0,0,3,1) \\
$588=294^++294^-$ & \begin{picture}(8,8)
\multiput(0,0)(0,-2){4}{\framebox (2,2)}
\multiput(2,0)(0,-2){4}{\framebox (2,2)}
\end{picture} & (0,0,0,4)+(0,0,4,0) \\
&&\\
\hline
    \end{tabular}
\end{table}

\thispagestyle{empty}
\setlength{\unitlength}{0.6mm}
{\small
\begin{table}[h]
\caption{ Massive excitations on $AdS_5$; 
$Sp(4) \times Sp(8)$ classification and Young tableaux}
\label{4dclass}
\begin{tabular}{ccccc}
\hline
Spin & $Sp(4)$ & dim & Sp(8) & dim \\
\hline
4 & 
\begin{picture}(12,8)
\multiput(0,0)(3,0){4}{\framebox (3,3)}
\multiput(0,-3)(3,0){4}{\framebox (3,3)}
\end{picture}
 & 55 & \bfone & 1 \\
&&&&\\
\hline
$\frac{7}{2}$ & 
\begin{picture}(12,8)
\multiput(0,0)(3,0){4}{\framebox (3,3)}
\multiput(0,-3)(3,0){3}{\framebox (3,3)}
\end{picture}
 & 80 &
\begin{picture}(12,8)
\put(0,0){\framebox (3,3)}
\end{picture}
  & 8 \\
&&&&\\
\hline
3 & 
\begin{picture}(12,8)
\multiput(0,0)(3,0){4}{\framebox (3,3)}
\multiput(0,-3)(3,0){2}{\framebox (3,3)}
\end{picture}
 & 81 &
\begin{picture}(12,8)
\multiput(0,0)(0,-3){2}{\framebox (3,3)}
\end{picture}
  & 27 \\
 & 
\begin{picture}(12,8)
\multiput(0,0)(3,0){3}{\framebox (3,3)}
\multiput(0,-3)(3,0){3}{\framebox (3,3)}
\end{picture}
 & 30 &
\begin{picture}(12,8)
\multiput(0,0)(3,0){2}{\framebox (3,3)}
\end{picture}
  & 36 \\
&&&&\\
\hline
$\frac{5}{2}$ & 
\begin{picture}(12,8)
\multiput(0,0)(3,0){4}{\framebox (3,3)}
\put(0,-3){\framebox (3,3)}
\end{picture}
 & 64 &
\begin{picture}(12,8)
\multiput(0,0)(0,-3){3}{\framebox (3,3)}
\end{picture}
  & 48 \\
&&&&\\
 & 
\begin{picture}(12,8)
\multiput(0,0)(3,0){3}{\framebox (3,3)}
\multiput(0,-3)(3,0){2}{\framebox (3,3)}
\end{picture}
 & 40 &
\begin{picture}(12,8)
\multiput(0,0)(3,0){2}{\framebox (3,3)}
\put(0,-3){\framebox (3,3)}
\end{picture}
  & 160\\
&&&& \\
\hline
2  & 
\begin{picture}(12,8)
\multiput(0,0)(3,0){4}{\framebox (3,3)}
\end{picture}
 & 35 &
\begin{picture}(12,8)
\multiput(0,0)(0,-3){4}{\framebox (3,3)}
\end{picture}
  & 42\\
&&&& \\
 & 
\begin{picture}(12,8)
\multiput(0,0)(3,0){3}{\framebox (3,3)}
\put(0,-3){\framebox (3,3)}
\end{picture}
 & 35' &
\begin{picture}(12,8)
\multiput(0,0)(0,-3){3}{\framebox (3,3)}\put(3,0){\framebox (3,3)}
\end{picture}
  & 315\\
&&&& \\
& 
\begin{picture}(12,8)
\multiput(0,0)(3,0){2}{\framebox (3,3)}
\multiput(0,-3)(3,0){2}{\framebox (3,3)}
\end{picture}
 & 14 &
\begin{picture}(12,8)
\multiput(0,0)(3,0){2}{\framebox (3,3)}
\multiput(0,-3)(3,0){2}{\framebox (3,3)}
\end{picture}
  & 308\\
&&&& \\
\hline
$\frac{3}{2}$& 
\begin{picture}(12,8)
\multiput(0,0)(3,0){3}{\framebox (3,3)}
\end{picture}
 & 20 &
\begin{picture}(12,8)
\multiput(0,0)(0,-3){4}{\framebox (3,3)}\put(3,0){\framebox (3,3)}
\end{picture}
  & 288\\
&&&& \\
& \begin{picture}(12,8)
\multiput(0,0)(0,-3){2}{\framebox (3,3)}\put(3,0){\framebox (3,3)}
\end{picture}
& 16 
& \begin{picture}(12,8)
\multiput(0,0)(0,-3){3}{\framebox (3,3)}\multiput(3,0)(0,-3){2}{\framebox (3,3)}
\end{picture}
& 792 \\
&&&& \\
\hline
1 & 
\begin{picture}(12,8)
\multiput(0,0)(3,0){2}{\framebox (3,3)}
\end{picture}
 & 10 &
\begin{picture}(12,8)
\multiput(0,0)(0,-3){4}{\framebox (3,3)}\multiput(3,0)(0,-3){2}{\framebox (3,3)}
\end{picture}
  & 792'\\
&&&& \\
&
\begin{picture}(12,8)
\multiput(0,0)(0,-3){2}{\framebox (3,3)}
\end{picture}
& 5 &
\begin{picture}(12,8)
\multiput(0,0)(0,-3){3}{\framebox (3,3)}\multiput(3,0)(0,-3){3}{\framebox (3,3)}
\end{picture}
  & 825\\
&&&& \\
\hline
$\frac{1}{2}$ & 
\begin{picture}(12,8)
\put(0,0){\framebox (3,3)}
\end{picture}
& 4 &
\begin{picture}(12,8)
\multiput(0,0)(0,-3){4}{\framebox (3,3)}\multiput(3,0)(0,-3){3}{\framebox (3,3)}
\end{picture}
  & 1056\\
&&&& \\
\hline
0 & 
 \bfone & 1 &
\begin{picture}(12,8)
\multiput(0,0)(0,-3){4}{\framebox (3,3)}\multiput(3,0)(0,-3){4}{\framebox (3,3)}
\end{picture}
  & 594\\
&&&& \\
\hline
\end{tabular}
\end{table}
}


\begin{table}[h]
 \caption{$Sp(4) \to SU(2) \times SU(2)$ decomposition --- spin max $(2,2)$}
\label{sp4dec}
    \begin{tabular}{ll}
\hline
dim $[Sp(4)]$ & $(SU(2),SU(2))$ rep \\
\hline
1 & $(0,0)$\\
&\\
4 &$ (\frac{1}{2} ,0) + (0,\frac{1}{2} )$ \\
&\\
5 & $(\frac{1}{2} , \frac{1}{2} )+ (0,0) $  \\
&\\
10 & $(1,0) + (\frac{1}{2} , \frac{1}{2} )+ (0,1) $ \\
&\\
16 & $(1,\frac{1}{2} ) + (\frac{1}{2} , 1) + (\frac{1}{2} ,0) + (0,\frac{1}{2} )$ \\
&\\
20 & $ (\frac{3}{2}, 0) +(0,\frac{3}{2}) + (1,\frac{1}{2} ) + (\frac{1}{2} ,1)$ \\
&\\
14 & $(1,1) + (\frac{1}{2} , \frac{1}{2} ) + (0,0)$ \\
&\\
35' & $(\frac{3}{2}, \frac{1}{2} ) + (\frac{1}{2} , \frac{3}{2})+ (1,1) + (1,0) + (0,1) + (\frac{1}{2} , \frac{1}{2} )$\\
&\\
35 & $ (2,0) + (0,2) + (\frac{3}{2}, \frac{1}{2} ) + (\frac{1}{2} , \frac{3}{2})+ (1,1) $\\
&\\
40 & $(\frac{3}{2},1) + (1,\frac{3}{2}) + (1,\frac{1}{2} )+(\frac{1}{2} ,1)+(\frac{1}{2} ,0)+(0,\frac{1}{2} )$\\
&\\
64 & $(2,\frac{1}{2} )+ (\frac{1}{2} , 2) +(\frac{3}{2},1) + (1,\frac{3}{2}) +
      (\frac{3}{2}, 0) +(0,\frac{3}{2}) + (1,\frac{1}{2} ) + (\frac{1}{2} ,1)$ \\
&\\
30 & $ (\frac{3}{2},\frac{3}{2})+ (1,1)+(\frac{1}{2} ,\frac{1}{2} )+(0,0)$\\
&\\
81 & $(2,1) + (1,2) + (\frac{3}{2},\frac{3}{2})+ (\frac{3}{2},\frac{1}{2})+(\frac{1}{2},\frac{3}{2})+
     (1,1)+ (1,0)+(0,1) +(\frac{1}{2} ,\frac{1}{2} )$\\
&\\
80 & $(2,\frac{3}{2}) + (\frac{3}{2},2)+(\frac{3}{2},1) +(1,\frac{3}{2})+(1,\frac{1}{2} )+(\frac{1}{2} ,1)
     +(\frac{1}{2} ,0)+(0,\frac{1}{2} )$\\ 
&\\
55 & $(2,2) + (\frac{3}{2},\frac{3}{2})+ (1,1)+(\frac{1}{2} ,\frac{1}{2} )+(0,0)$\\
&\\
\hline
    \end{tabular}
\end{table}

\begin{table}[h]
 \caption{$Sp(8) \to SU(4)$}
\label{sp8dec}
    \begin{tabular}{ll}
\hline
dim $[Sp(8)]$ & dim $[SU(4)]$ \\
\hline
1 & 1 \\
&\\
 8 & $4+ \bar 4$ \\
&\\
27 & 6+6+15 \\
&\\
36 & $ 1+10 + {\bar {10}}+15 $\\
&\\
48 & $4 + \bar 4 + 20 + {\bar {20}}$ \\
&\\
160 & $4 + \bar 4 + 20 + {\bar {20}}+ 20 + {\bar {20}} + 36 + {\bar {36}} $\\
&\\
120 & $4 + \bar 4 + 20'' + {\bar {20}}'' + 36 + {\bar {36}} $\\
&\\
42 &  $ 1 + 1+10 + {\bar {10}}+20' $\\
&\\
315 & $6+6+10 + {\bar {10}}+ 15+15+15 + 20' + 45 + {\bar {45}}+64+64$\\
&\\
308 & $1+10 + {\bar {10}}+15 +20' +20'+20'+64+64+84 $\\
&\\
288 & $ 4 + {\bar 4}+ 4 + {\bar 4}+20 + {\bar {20}}+ 20'' + {\bar {20}}''+36 +{\bar{36}}+ 60 +{\bar{60}}$\\
&\\
792 & $ 4 + {\bar 4}+ 20 + {\bar {20}}+20 + {\bar {20}}+ 20 + {\bar {20}}+36 +{\bar{36}}+36 +{\bar{36}}$\\
    &    $+ 60 +{\bar{60}} + 60 +{\bar{60}}  + 140 +{\bar{140}} $\\
&\\
792' & $6+6+6+6+ 15+15+15+45+{\bar{45}}+45+{\bar{45}}+50+50$\\
     & $+64+64+70+{\bar{70}}+175$\\
&\\
825 & $1+10+{\bar{10}}+10+{\bar{10}}+15+20'+20'+45+{\bar{45}}$\\
    & $+64+64+84+126+{\bar{126}}+175$ \\
&\\
1056 & $4 + {\bar 4}+ 4 + {\bar 4}+20 + {\bar {20}}+20 + {\bar {20}}+ 20'' + {\bar {20}}''+36 +{\bar{36}}$\\
    &    $+ 60 +{\bar{60}} + 84' +{\bar{84}}'  + 140 +{\bar{140}}+ 140' +{\bar{140}}' $\\
&\\
594 & $1+1+1 + 10 + {\bar{10}}+ 10 + {\bar{10}}+20'+20'+35+{\bar{35}}$\\
    & $+84+105+126+{\bar{126}}$\\
\hline
    \end{tabular}
\end{table}


\thispagestyle{empty}
\setlength{\unitlength}{0.8mm}
{\small
\begin{table}[h]
 \caption{$SU(4)$ reps of interest}
\label{su4rep}
    \begin{tabular}{lcl}
\hline
rep & Young tableau  & Dynkin label \\
\hline
1 & \begin{picture}(8,8)
\multiput(0,0)(0,-2){4}{\framebox (2,2)}
\end{picture}
  &  (0,0,0) \\
4 & \begin{picture}(8,8)
\put(0,0){\framebox (2,2)}
\end{picture} & (1,0,0) \\
$\bar 4$ &\begin{picture}(8,8)
\multiput(0,0)(0,-2){3}{\framebox (2,2)}
\end{picture} &  (0,0,1) \\
6 &\begin{picture}(8,8)
\multiput(0,0)(0,-2){2}{\framebox (2,2)}
\end{picture} &  (0,1,0)) \\
10 &\begin{picture}(8,8)
\multiput(0,0)(2,0){2}{\framebox (2,2)}
\end{picture} &  (2,0,0)) \\
20 &\begin{picture}(8,8)
\multiput(0,0)(0,-2){2}{\framebox (2,2)}
\put(2,0){\framebox (2,2)}
\end{picture} &  (1,1,0) \\
20'' &\begin{picture}(8,8)
\multiput(0,0)(2,0){3}{\framebox (2,2)}
\end{picture} &  (3,0,0) \\
15 &\begin{picture}(8,8)
\multiput(0,0)(0,-2){3}{\framebox (2,2)}
\put(2,0){\framebox (2,2)}
\end{picture} &  (1,0,1)) \\
20' & \begin{picture}(8,8)
\multiput(0,0)(2,0){2}{\framebox (2,2)}
\multiput(0,-2)(2,0){2}{\framebox (2,2)}
\end{picture} & (0,2,0) \\
45 &\begin{picture}(8,8)
\multiput(0,0)(2,0){3}{\framebox (2,2)}
\put(0,-2){\framebox (2,2)}
\end{picture} &  (2,1,0) \\
35 & \begin{picture}(8,8)
\multiput(0,0)(2,0){4}{\framebox (2,2)}
\end{picture} & (4,0,0) \\
36 & \begin{picture}(8,8)
\multiput(0,0)(2,0){3}{\framebox (2,2)}
\put(0,-2){\framebox (2,2)}
\put(0,-4){\framebox (2,2)}
\end{picture} & (2,0,1) \\
60 & \begin{picture}(8,8)
\multiput(0,0)(2,0){3}{\framebox (2,2)}
\multiput(0,-2)(2,0){2}{\framebox (2,2)}
\end{picture} & (1,2,0) \\
84' & \begin{picture}(8,8)
\multiput(0,0)(2,0){4}{\framebox (2,2)}
\put(0,-2){\framebox (2,2)}
\end{picture} & (3,1,0) \\
50 &\begin{picture}(8,8)
\multiput(0,0)(2,0){3}{\framebox (2,2)}
\multiput(0,-2)(2,0){3}{\framebox (2,2)}
\end{picture} &  (0,3,0) \\
126 &\begin{picture}(8,8)
\multiput(0,0)(2,0){4}{\framebox (2,2)}
\multiput(0,-2)(2,0){2}{\framebox (2,2)}
\end{picture} &  (2,2,0) \\
 64 & \begin{picture}(8,8)
\multiput(0,0)(2,0){3}{\framebox (2,2)}
\multiput(0,-2)(2,0){2}{\framebox (2,2)}
\put(0,-4){\framebox (2,2)}
\end{picture} &(1,1,1) \\
140' & \begin{picture}(8,8)
\multiput(0,0)(2,0){4}{\framebox (2,2)}
\multiput(0,-2)(2,0){3}{\framebox (2,2)}
\end{picture} & (1,3,0) \\
70 & \begin{picture}(8,8)
\multiput(0,0)(2,0){4}{\framebox (2,2)}
\put(0,-2){\framebox (2,2)}
\put(0,-4){\framebox (2,2)}
\end{picture} & (3,0,1) \\
140 & \begin{picture}(8,8)
\multiput(0,0)(2,0){4}{\framebox (2,2)}
\multiput(0,-2)(2,0){2}{\framebox (2,2)}
\put(0,-4){\framebox (2,2)}
\end{picture} &  (2,1,1) \\
84 & \begin{picture}(8,8)
\multiput(0,0)(2,0){4}{\framebox (2,2)}
\multiput(0,-2)(2,0){2}{\framebox (2,2)}
\multiput(0,-4)(2,0){2}{\framebox (2,2)}
\end{picture} & (2,0,2) \\
105 & \begin{picture}(8,8)
\multiput(0,0)(2,0){4}{\framebox (2,2)}
\multiput(0,-2)(2,0){4}{\framebox (2,2)}
\end{picture} & (0,4,0) \\
175 & \begin{picture}(8,8)
\multiput(0,0)(2,0){4}{\framebox (2,2)}
\multiput(0,-2)(2,0){3}{\framebox (2,2)}
\put(0,-4){\framebox (2,2)}
\end{picture} & (1,2,1) \\
&&\\
\hline
    \end{tabular}
\end{table}
}

\end{document}